\documentclass[prl,aps,showpacs,twocolumn,superscriptaddress]{revtex4}
\usepackage{times,mathptm}
\usepackage{amsmath,amsfonts,bm}
\usepackage[dvips]{graphicx}

\begin{document}


\title{Diagnosis of weaknesses in modern error correction codes: a
  physics approach}

\author{M.G.~Stepanov}
\affiliation{Theoretical Division and Center for Nonlinear Studies, Los
  Alamos National Laboratory, Los Alamos, NM 87545, USA}
\affiliation{Institute of Automation and Electrometry, Novosibirsk
  630090, Russia}
\author{V.~Chernyak}
\affiliation{Department of Chemistry, Wayne State University, 5101 Cass
  Avenue, Detroit, MI 48202, USA}
\author{M.~Chertkov}
\affiliation{Theoretical Division and Center for Nonlinear Studies, Los
  Alamos National Laboratory, Los Alamos, NM 87545, USA}
\author{B.~Vasic}
\affiliation{Department of Electrical and Computer Engineering and
  Department of Mathematics, University of Arizona, Tucson, AZ 85721,
  USA}


\begin{abstract}
One of the main obstacles to the wider use of the modern error-correction codes is that, due to the
complex behavior of their decoding algorithms, no systematic method which would allow
characterization of the Bit-Error-Rate (BER) is known. This is especially true at the weak noise
where many systems operate and where coding performance is difficult to estimate because of the
diminishingly small number of errors. We show how the instanton method of physics allows one to
solve the problem of BER analysis in the weak noise range by recasting it as a computationally
tractable minimization problem.
\end{abstract}

\pacs{89.70.+c, 02.50.-r}

\maketitle

Modern technologies, as well as many natural and sociological systems, rely heavily on a wide
range of error-correction mechanisms to compensate for their inherent unreliability and to ensure
faithful transmission, processing and storage of information. There has been a great deal of
research activity in coding theory in the last half a century that has culminated in the recent
discovery of coding schemes \cite{63Gal,93BGT,99Mac} that approach a reliability limit set by
classical information theory \cite{48Sha}. The problem considered in this paper is of a special
interest because of a unique feature of the modern coding schemes, which is referred to as an
error floor \cite{03MP,04Ric}. Error floor is a phenomenon characterized by an abrupt degradation
of the coding scheme performance, as measured by the BER, from the so-called water-fall regime of
moderate Signal-to-Noise Ratio (SNR) to the absolutely different error-floor asymptotic achieved
at high SNR.  To estimate the error-floor asymptotic in the modern high-quality systems is a
notoriously difficult task. Typical required BER values are $10^{-12}$ for an optical
communication system, $10^{-15}$ for hard drive systems in personal computers and as small as
$10^{-20}$ for storage systems used in banks and financial institutions. However, direct numerical
methods, e.g.\ Monte Carlo, cannot be used to determine BER below $10^{-9}$.

To address this challenge we suggest a physics-inspired approach that
ultimately solves the problem of the error-floor analysis. The method is
coined the ``instanton" method, after a theoretical particle in quantum
physics that lasts for only an instant, occupying a localized portion of
space-time \cite{75BPST}. Statistical physics uses the word instanton to
describe a microscopic configuration which, in spite of its rare
occurrence, contributes most to the macroscopic behavior of the system
\cite{65Lif}. Our instanton is the most probable configuration of the
noise to cause a decoding error.

We consider a model of a general communication system with error correction \cite{48Sha}. Data
originating from an information source are parsed into fixed length words. Each word is encoded
into a longer codeword and transmitted through a noisy channel (e.g., radio or optical link,
magnetic or optical data storage system, etc.). The decoder tries to reconstruct the original
codeword using the knowledge of the noise statistics and the structure of the code. Error
resilience is achieved at the expense of introduced redundancy, and information theory gives
conditions for the existence of finite redundancy error correction codes. However it does not give
a method for realizing decoders of low complexity. In general there is no better way to
reconstruct the codeword that was most likely transmitted than to compare the likelihoods of all
possible codewords. However, this Maximal Likelihood (ML) algorithm becomes intractable already
for codewords that are tens of bits long.

A novel exciting era has started in coding theory with the discovery of Low-Density Parity-Check
(LDPC) \cite{63Gal,99Mac,03RU,04BV} and turbo \cite{93BGT} codes. These codes are special, not
only because they can approach very close to the virtually error-free transmission limit, but
mainly because a computationally efficient, so-called iterative, decoding scheme is readily
available. When operating at moderate noise values these approximate decoding algorithms show an
unprecedented ability to correct errors, a remarkable feature that has attracted a lot of
theoretical attention \cite{95WLK,96Wib,99FKKR,01WF,01KFL,03MP,04Ric}. (Notice also an alternative
statistical physics inspired approach \cite{89Sou} that offered an important insight into the
extraordinary performance of the iterative decoding \cite{01Mon,02YFW,03Mez}.) It is believed that
the error floor is a fundamental consequence of iterative decoding, and that the approximate
algorithms mentioned above are incapable of matching the performance of ML decoding beyond the
error-floor threshold. The importance of error-floor analysis was recognized in the early stages
of the turbo codes revolution \cite{96BM}, and it soon became apparent that LDPC codes are also
not immune from the error-floor deficiency \cite{01MB,03TJVW,04Ric}. The main approaches to the
error-floor analysis problem proposed to date include: (i) a heuristic approach of the importance
sampling type \cite{04Ric}, utilizing theoretical considerations developed for a typical randomly
constructed LDPC code performing over the very special binary-erasure channel \cite{02DPTRU}, and
(ii) deriving lower bounds for BER \cite{04VK}.

Our approach to the error-floor analysis is different: we suggest an
efficient numerical scheme, which is ab-initio by construction,
i.e.\ the scheme requires no additional assumptions (e.g.\ no sampling).
The numerical scheme is also accurate at producing configurations whose
validity, as of actual optimal noise configurations, can be verified
theoretically. Finally, the instanton scheme is also generic, in that
there are no restrictions related to the channel or decoding.

{\bf Error-correction scheme.} A message word consisting of $K$ bits is encoded in an $N$-bit long
codeword, $N>K$. In the case of binary, linear coding, a convenient representation of the code is
given by $M\geq N-K$ constraints, often called parity checks or simply checks. Formally,
${\bm\sigma} = (\sigma_{1}, \dots, \sigma_{N})$ with $\sigma_{i} = \pm 1$, is one of the $2^{K}$
codewords if and only if $\prod_{i\in \alpha } \sigma_{i} = 1$ for all checks $\alpha = 1, \dots,
M$, where $i\in\alpha$ if the bit $i$ contributes the check $\alpha$. The relation between bits and
checks (we use $i\in\alpha$ and $\alpha\ni i$ interchangeably) is often described in terms of the
$M\times N$ parity-check matrix $\hat{ H}$ consisting of ones and zeros: $H_{\alpha i}=1$ if
$i\in\alpha$ and $H_{\alpha i}=0$ otherwise. A bipartite graph representation of $\hat{H}$, with
bits marked as circles checks marked as squares and edges corresponding to respective nonzero
elements of $\hat{H}$, is usually called Tanner graph of the code. For an LDPC code $\hat{H}$ is
sparse, i.e.\ most of the entries are zeros. Transmitted through a noisy channel, a codeword gets
corrupted due to the channel noise, so that the channel output (receiver) is ${\bm
x}\neq{\bm\sigma}$. Even though an information about the original codeword is lost at the
receiver, one still possesses the full probabilistic information about the channel, i.e.\ the
conditional probability, $P({\bm x}|{\bm \sigma}')$, for a codeword ${\bm \sigma}'$ to be a
preimage for the output word ${\bm x}$, is known. In the case of independent noise samples the
full conditional probability can be decomposed into the product, $P({\bm x}|{\bm \sigma}')=\prod_i
p(x_i|\sigma_i')$. A convenient characteristic of the channel output at a bit is the so-called
log-likelihood, $h_i=\log[p(x_i|+1)/p(x_i|-1)]/2s^2$, measured in the units of the SNR squared,
$s^2$. (In the physics formulation \cite{89Sou,01Mon,02YFW,03Mez} ${\bm h}$ is called the magnetic
field.) The decoding goal is to infer the original message from the received output ${\bm x}$. ML
decoding (which generally requires an exponentially large number, $2^K$, of steps) corresponds to
finding the most probable transmitted codeword given ${\bm x}$. Belief Propagation (BP) decoding
\cite{63Gal,99Mac,03RU,03Mez} constitutes a fast (linear in $K, N$) yet generally approximate
alternative to ML. As shown in \cite{63Gal} the set of equations describing BP becomes exactly
equivalent to the so-called symbol Maximum-A-Posteriori (MAP) decoding in the loop-free
approximation (a similar construction in physics is known as the Bethe-tree approximation
\cite{35Bet}), while in the low-noise limit, $s\to\infty$, ML and MAP become indistinguishable and
the BP algorithm reduces to the min-sum algorithm:
\begin{equation}
  \eta^{(n+1)}_{i\alpha} = h_i + \sum_{\beta\neq\alpha}^{\beta\ni i}
  \prod_{j\neq i}^{j\in\beta} \mbox{sign}
    \big[ \eta^{(n)}_{j\beta} \big] \min_{j\neq i}^{j\in\beta} \big|
    \eta^{(n)}_{j\beta} \big|,
  \label{min_sum}
\end{equation}
where the message field $\eta^{(n)}_{i\alpha}$ is defined on the edge that connects bit $i$ and
check $\alpha$ at the $n$-th step of the iterative procedure and $\eta^{(0)}_{i\alpha} \equiv 0$.
The result of decoding is determined by magnetizations, $m^{(n)}_i$, defined by the right-hand-side
of Eq.~(\ref{min_sum}) with the restriction $\beta\neq\alpha$ dropped.  The BER at a given bit $i$
becomes
\begin{equation}
  B_{i}=\int d{\bm x}\ \theta\!\left(-m_{i}\{\bm x\} \right)
    P({\bm x}|{\bm 1}),
  \label{BER}
\end{equation}
where $\theta(z)=1$ if $z>0$ and $\theta(z)=0$ otherwise;
${\bm\sigma} = {\bm 1}$ is assumed for the input (since in a
symmetric channel the BER is invariant with respect to the choice
of the input codeword).

When the BER is small, the integral over output configurations
${\bm x}$ in Eq.~(\ref{BER}) is approximated by, $B_i\sim P({\bm
x}_{\rm inst}|{\bm 1})$, where ${\bm x}_{\rm inst}$ is the special
instanton configuration of the output minimizing $P({\bm x}|{\bm
1})$ under the error-surface condition, $m_i\{{\bm x}\}=0$. For
the common model of the white symmetric Gaussian channel,
$p(x|\sigma) = \exp(-s^2 (x - \sigma)^2/2)/\sqrt{2\pi/s^2}$,
finding the instanton, ${\bm \varphi}_{\rm inst} = {\bm 1} - {\bm
x}_{\rm inst} \equiv l({\bm u}){\bm u}$, turns into minimizing the
length $l({\bm u})$ with respect to the unit vector in the noise
space ${\bm u}$, where $l({\bm u})$ measures the distance from the
zero-noise point to the point on the error surface corresponding
to ${\bm u}$.

{\bf Finding the instanton numerically.}  In our numerical scheme,
the value of the length $l({\bm u})$ for any given unit vector
${\bm u}$ was found by the bisection method. The minimum of
$l({\bm u})$ was found by a downhill simplex method also called
``amoeba'' \cite{88Pre}, with accurately tailored (for better
convergence) annealing. The numerical instanton method was first
successfully verified in \cite{04CCSV} against analytical
loop-free results.

Our demonstrative example is the $(155,64,20)$ LDPC code described in \cite{01TSF}. (The parity
check matrix of the code is shown in Fig.~S1 of Appendix A.) The code includes $155$ bits and $93$
checks. Each bit is connected to three checks while any check is connected to five bits. The
minimal Hamming distance of the code is $l_{\rm ML}^2 = 20$, i.e.\ at $s\gg 1$, and if the
decoding is ML, BER becomes $\sim \exp(-20\cdot s^2/2)$. (See Fig.~S2 of Appendix A for Monte Carlo
evaluation of BER vs SNR for the code.) We aim to find and describe the instanton(s) that
determines BER in the error-floor regime (for min-sum decoding): $\sim \exp(-l_{\rm ef}^2 \cdot
s^2/2)$ with $l_{\rm ef}^2 < l_{\rm ML}^2 = 20$. Our numerical, and subsequent theoretical,
analyses suggest that the instantons, as well as $l_{\rm ef}$, do depend on the number of
iterations. We do not detail this rich dependence here, focusing primarily on the already
nontrivial case of four iterations.

The instanton with the minimal length of $ l_{\sf a}^2 = 46^2/210 \approx 10.076$ is shown in the
upper part of Fig.~1A, see also Fig.~S3 of Appendix A. Everywhere away from the $12$-bit pattern
the noise is numerical zero. The resulting nonzero noise values are proportional to integers
(within numerical precision). If decoding starts from the instanton configuration of the noise,
magnetization is exactly zero at the bit number ``77''. This minimal length instanton controls BER
at $s\to\infty$, however, for any large but finite $s$ one should also account for many other
"close" instantons with $l({\bm u}) \approx l_{\sf a}$, thus approximating $B_i\sim\sum_{\rm
inst}P({\bm x}_{\rm inst}|{\bm 1})$. Two instanton configurations shown in Fig.~1B and Fig.~1C
represent two local minima $l_{\sf b}^2 = 806/79 \approx 10.203$ and $l_{\sf c}^2 = 44^2/188
\approx 10.298$ respectively, that are the closest to the minimal one. (See also Appendix A
Figs.~S4--S5.) These instantons were found as a result of multiple attempts at ``amoeba"
minimization.

\begin{figure*}
 \centerline{\includegraphics[width=4.9in]{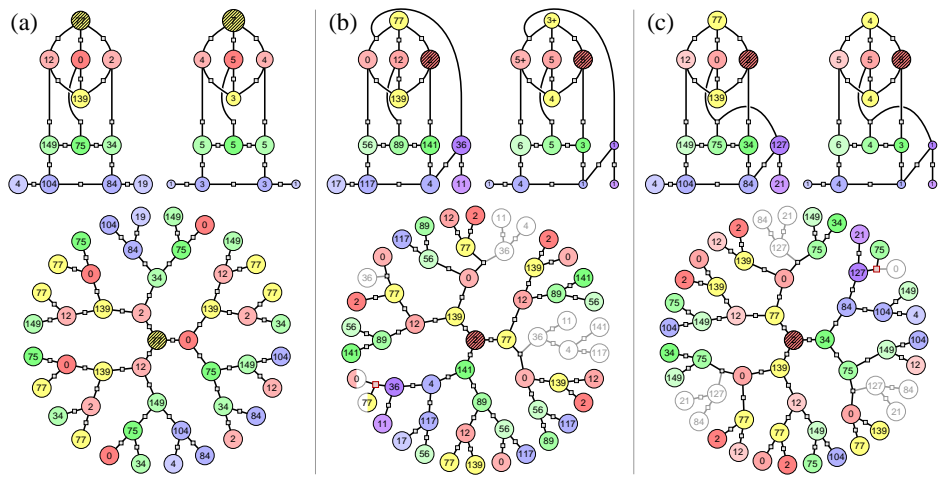}}
  \caption{{\bf Parts of the full Tanner graph with nonzero noise for
    the instantons}, corresponding to (a) simple, (b) degenerate and (c)
    sign-alternating pseudo-codewords, are shown in three panels each
    consisting of three diagrams. Bits are numbered according to the
    $(155,64,20)$-code definition (top left and bottom) and the noise
    level (top right, where the area of a bit/circle is proportional to
    the corresponding number). For the computational tree (bottom panel)
    the bits drawn in color participate in the pseudo-codeword and the
    shaded bit marks the error position. The marked checks/squares
    correspond to the points of (b) degeneracy and of (c)
    sign-alternation. }
  \label{Fig:1}
\end{figure*}

\begin{figure*}
 \centerline{\includegraphics[width=4.9in]{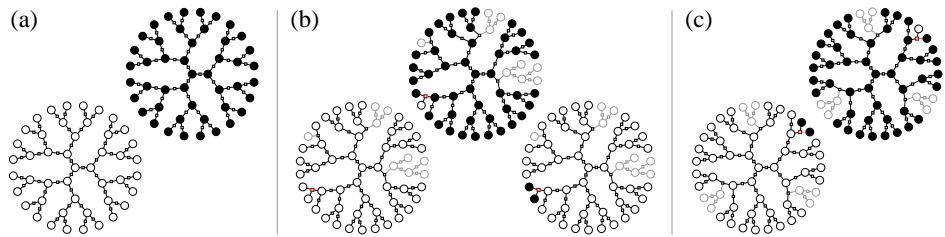}}
  \caption{{\bf Interpretation of the instanton as a median within a set
    of pseudo-codewords.} Three panels show the set of pseudo-codewords
    for the three instantons described in Fig.~1. Bits on the
    computational tree painted in white/black correspond to $+1/-1$.
    Other notations/marks are in accordance with the captions of
    Fig.~1.}
  \label{Fig:2}
\end{figure*}

{\bf Interpretation of the instantons found.} The remarkable integer/rational structure of the
instantons found numerically by ``amoeba" admits a theoretical explanation. Our algebraic
construction generalizes the computational tree approach of Wiberg \cite{96Wib}. The computational
tree is built by unwrapping the Tanner graph of a given code into a tree from a bit for which we
would like to determine the probability of error. (The erroneous bit is shaded in Fig.~1.) The
number of generations in the tree is equal to the number of BP iterations (for more details see
\cite{95WLK}). As observed in \cite{96Wib}, the result of decoding at the shaded bit of the
original code is exactly equal to the decoding result in the tree center. It should be noted that
once magnetic fields representing an instanton are distributed on the tree, one can verify
directly (by propagating messages from the leaves to the tree center) that the algorithm produces
zero magnetization at the tree center. Any check node processes messages coming from the tree
periphery in the following way: (i) the message with the smallest absolute value (we assume no
degeneracy in the beginning) is passed, (ii) the source bit of the smallest message is colored,
and (iii) the sign of the product of inputs is assigned to the outcome. At any bit that lies on
the colored leaves-to-center path the incoming messages are summed up. The initial messages at any
bit of the tree are magnetic fields and, therefore, the result obtained in the tree center is a
linear combination of the magnetic fields with integer coefficients. The integer $n_i$
corresponding to bit $i$ of the original graph is the sum of the signatures over all colored
replicas of $i$ on the computational tree. Therefore, the condition at the tree center becomes
$\sum_i n_i h_i = 0$. Returning to the original graph and maximizing the integrand of
Eq.~(\ref{BER}) with the condition enforced we arrive at the following expressions for the
instanton configuration and the effective weight, respectively:
\begin{equation}
  \varphi_j \!=\! n_j \Big( \sum_i n_i \Big)
    \Big/ \Big( \sum_i n_i^2 \Big),
    \quad l^2\! =\! \Big( \sum_i n_i \Big)^2 \Big/ \Big( \sum_i n_i^2
    \Big) ,
  \label{Wib}
\end{equation}
where the equation applies to the Gaussian channel, however its generalizations to any other
channel is straightforward. One can check directly (e.g.\ looking at Fig.~1A) that
Eqs.~(\ref{Wib}) are satisfied for the minimum weight instanton. In this case we find that the
signature of any colored message before and after processing through a check remains intact, and
thus the resulting $n_i$ for any colored bit is just a total number of the bit's replicas. The
structure of this instanton is exactly equivalent to one of the codewords on the computational
tree, called a pseudo-codeword as generically it does not correspond to a codeword on the original
graph \cite{96Wib}. However, Eq.~(\ref{Wib}) also suggests another possibility that goes beyond the
standard pseudo-codeword construction \cite{96Wib}. In the case shown in Fig.~1C the colored part
of the tree does correspond to a pseudo-codeword (by structure), however the pale part of the
computational tree cannot be neglected as the noise values at these nodes are nonzero. This
peculiarity is due to the fact that some of the checks shown in the upper part of Fig.~1C are
connected to more than two colored bits. One finds that the signature of the message propagating
from bit ``75'' to bit ``127'' alternates because of the pale ``0'' lying on a leaf,
$8/47=|h_{0}|>|h_{75}|=3/47$, $h_{0} = -8/47 <0$. This modifies $n_{75}$ making it equal to $4$,
as one of the $6$ replicas of the bit ``75'' contributes to the total count $-1$ instead of $+1$.
Moreover, looking at Fig.~1B one finds that the instanton can be even more elaborate as the number
of replicas for some bits becomes fractional. (``+''-sign on Fig.~1B corresponds to $+7/18$.) This
is actually the degenerate case with a colored structure bifurcating at a check (connected to the
bits ``0'',``77'' and ``36'') so that the messages entering the check from two distinct periphery
have different signatures but are exactly equal to each other by the absolute value, $h_{0} =
-h_{77}=18/79$. Eq.~(\ref{Wib}) does not work for this case, but the following generalization
corrects the problem: one needs to introduce an additional condition accounting for the
degeneracy. In our example this extra condition can be simply stated as $h_{0}= -h_{77}$. (See
Appendix A Fig.~S6.)

Instantons also allow for a complementary interpretation. A decoding error occurs when the
magnetization in the computational tree center, which can be considered as a sum over all
pseudo-code words weighted by, $\exp\left(s^2\sum_{i}h_{i}p_{i}\right)$, turns to zero (with
$p_{i}$ being the number of bit $i$ replicas with $-1$ sign in the pseudo-codeword). In the case
of high SNR (large magnetic fields) the sum is dominated by the pseudo-codewords of maximal
weight. Therefore, any instanton, as a configuration of magnetic fields, should be equidistant
from some set of $k\ge2$ pseudo-codewords: $\sum_i h_ip^{(1)}_i=\dots=\sum_i h_i p^{(k)}_i$, where
at least one of them has $+1$ value and at least one has $-1$ value in the tree center to achieve
zero magnetization. And indeed the set of relevant pseudo-codewords for the $(155,64,20)$ code
example, shown in Fig.~2, is a pair in the cases (a), (c) and a triple
in the case (b). (See
Appendixes.)

To conclude, in this Letter we demonstrated that the instanton approach is a very powerful,
practical and generic instrument for quantitative analysis of the error floor. The success makes
us confident that this novel method will be indispensable for future design of good and practical
error-correcting schemes.

The authors acknowledge very useful and inspiring discussions with participants of workshop on
``Applications of Statistical Physics to Coding Theory'' sponsored by Los Alamos National
Laboratory that took place January 10--12, 2005 in Santa Fe, NM, USA. This work was supported in
part by DOE under LDRD program at Los Alamos National Laboratory, by the NSF under Grants
CCR-0208597 and ITR-0325979 and by the INSIC.

\appendix

\section{Appendix A: Supplementary Figures}
\vspace{.25in}

\centerline{\includegraphics[width=3.14in]{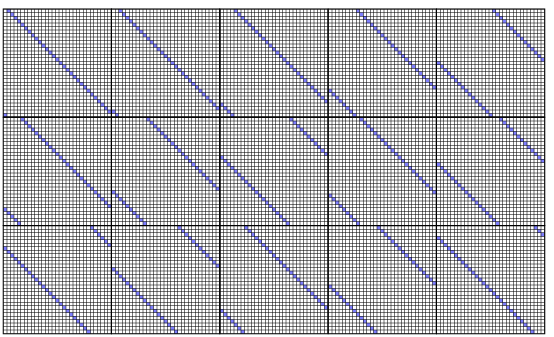}}
  \vspace{.025in}
  \noindent
  {\bf Figure S1.}
  {\it Parity check matrix $\hat{H}$ for the $(155,64,20)$ LDPC code.}
    The matrix consists of $3\times 5$ blocks. Each block is a square
    $31\times 31$ matrix. Empty/filled elements of the matrix stand for
    $0/1$. Bits are numbered from ``0'' to ``154''. The girth of this
    code is eight.

\vspace{.5in}

 \centerline{\includegraphics[width=3in]{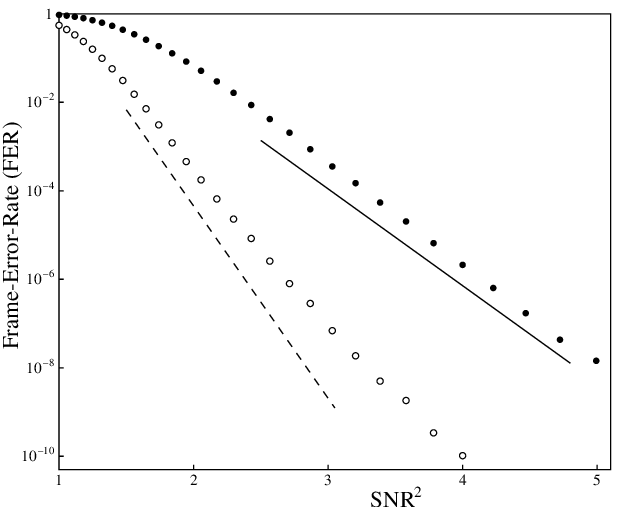}}
  \vspace{.025in}
  \noindent
  {\bf Figure S2.}
  {\it Frame-Error-Rate (FER) vs SNR$^2$ for the $(155,64,20)$ code and
    Belief Propagation decoding.} The filled/empty circle-marks
    correspond to result of Monte Carlo evaluation of FER for $4/1024$
    iterations of BP. The straight/dashed line corresponds to the
    (a)-instanton asymptotic, $\sim \exp[-(46^2/210)\cdot s^2/2]$
    and the ML asymptotic, $\sim\exp[-20\cdot s^2/2]$. \vfill

\vspace{.5in}

  \centerline{\includegraphics[width=3in]{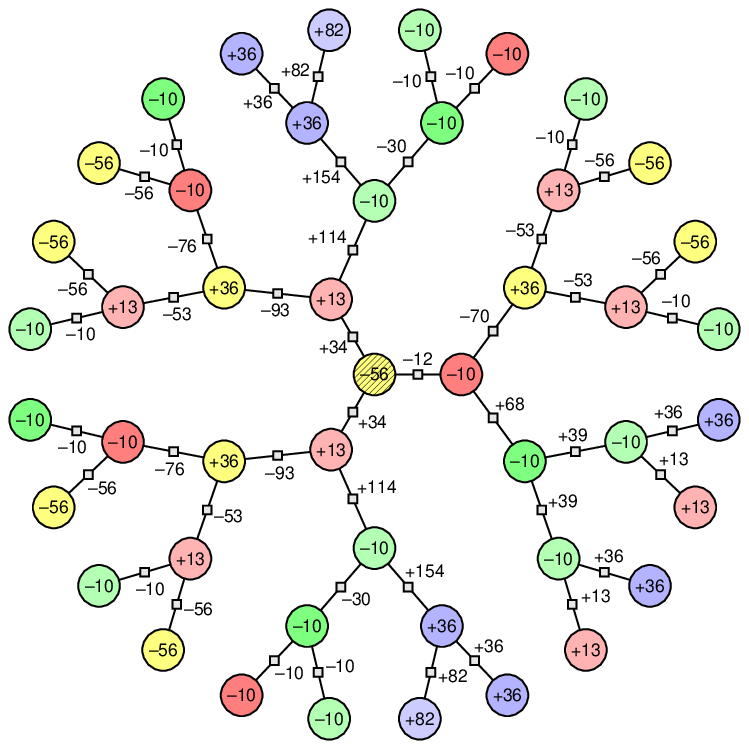}}
  \vspace{.125in}
  \noindent
  {\bf Figure S3.}
  {\it Min-sum decoding for the (a) instanton on the computational
tree.} All numbers are in units of $1/105$. The numbers shown on the bits/circles are the values
of the magnetic fields, $h$. The number shown next to each check/square is the message, $\eta$,
arriving at the check on the $q$-th step of the iteration procedure, where $4-q$ corresponds to
the number of bits separating this check from the tree center. Other notations/marks are in
accordance with the caption of Fig.~1 of the main text. \vfill

\vspace{.5in}

  \centerline{\includegraphics[width=3in]{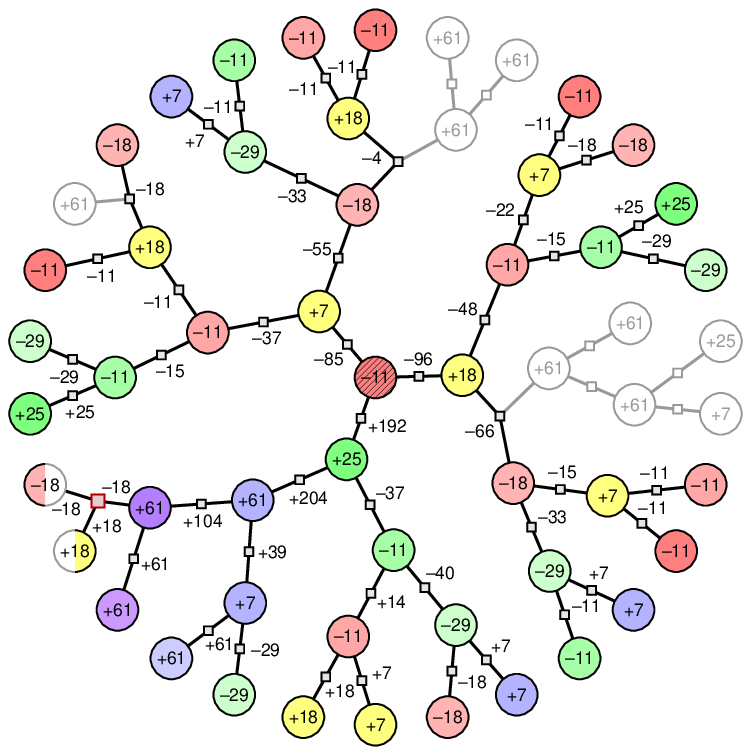}}
  \vspace{.125in}
  \noindent
  {\bf Figure S4.}
  {\it Min-sum decoding for the (b) instanton on the computational
tree.} All numbers are in units of $1/79$. Other notations/marks and explanations are in
accordance with captions of Fig.~1 of the main text and Fig.~S3.

\vspace{.5in}

  \centerline{\includegraphics[width=3in]{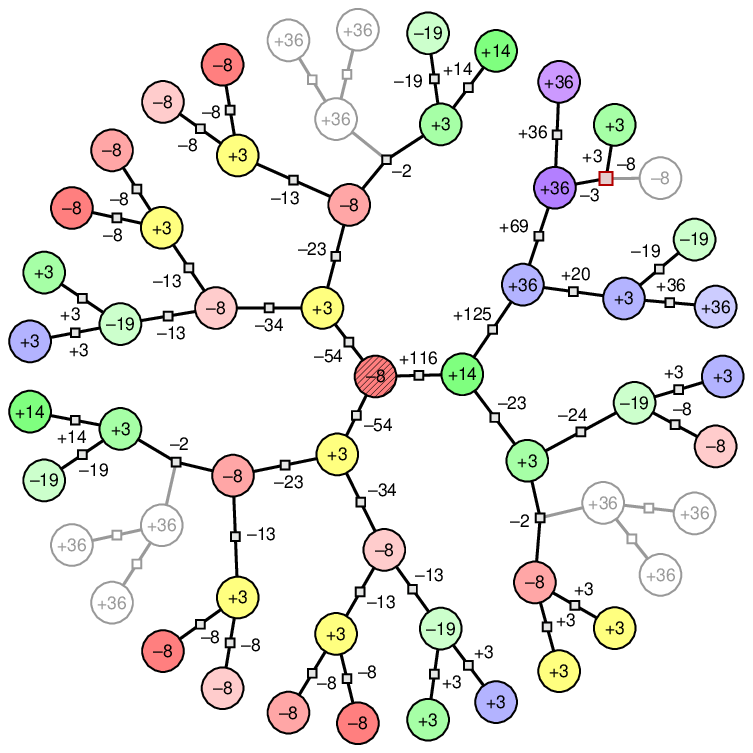}}
  \vspace{.125in}
  \noindent
  {\bf Figure S5.}
  {\it Min-sum decoding for the (c) instanton on the computational
tree.} All numbers are in the units of $1/47$. Other notations/marks and explanations are in
accordance with captions of Fig.~1 of the main text and Fig.~S3.

\vspace{.5in}

  \centerline{\includegraphics[width=3in]{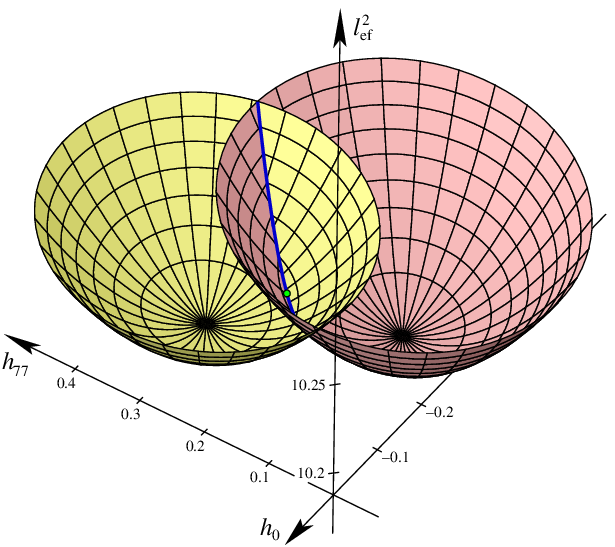}}
  \vspace{.125in}
  \noindent
  {\bf Figure S6.}
  {\it Geometrical interpretation of the degeneracy observed in case
(b).}
    The plot shows distance $l^2$ for the noise configuration in case
(b),
    minimized with respect to all the magnetic fields except $h_0$ and
    $h_{77}$, vs the two remaining fields. The relevant part of the
    $(h_0,h_{77})$ plane is the quadrant $h_0 < 0$, $h_{77} > 0$.
    Looking for the instanton in the $|h_0| < |h_{77}|$ domain and thus
    assuming that the ``0'' bit connected to the red check in Fig.~1B of
    the main text is colored while bit ``77'' is pale, one gets the paraboloid,
    shown in pink in the Figure, that achieves its minimum outside of
    the domain, i.e.\ in the $|h_0| > |h_{77}|$ semi-quadrant. Similarly, if
    one looks for the instanton in the $|h_0| > |h_{77}|$ domain one
    again finds an inconsistency as the respective paraboloid,  shown in
    yellow in the Figure achieves its minimum in the $|h_0| < |h_{77}|$
    semi-quadrant. Therefore, the point of actual minimum,
    shown by the green
    dot on the Figure, lies exactly at the minimum of the angled join of
    the two domains, $|h_0|=|h_{77}|$.

\section{Appendix B: Instantons for the min-sum decoding}

These notes consist of three parts. The first part is devoted to explaining how the entire
instanton family for an arbitrary LDPC code decoded by the min-sum algorithm can be fully
characterized using the computational tree approach.  The second part describes an alternative
exposition which allows one to represent an instanton as a configuration of magnetic fields
equidistant from some set of codewords on the computational tree. The third part formulates a
relation between the theoretical and numerical approaches and suggests challenges and questions
that need to be addressed in the future.

\subsection{Colored/signature structure and constraint minimization}

The basic object for our construction is the computational tree and its
col\-or\-ed/pale/un\-colored parts (as briefly introduced in the main text). The computational
tree is a tree constructed by a simple unwrapping of the Tanner graph of the code into a tree
starting from the bit where the BER is calculated. The number of generations on the tree is equal
to the number of iterations of the decoder, $n_{\rm it}$. If $n_{\rm it}$ is larger than, or equal
to, a quarter of the code girth (defined as the length of the shortest loop of the original Tanner
graph, measured by the number of edges within the loop), then the computational tree contains more
than one replica of some of the bits of the original code.

Consider an arbitrary configuration of magnetic field, ${\bm h}$ (or noise field, ${\bm\varphi}$)
on the Tanner graph, and on the computational tree respectively. Calculating the magnetization (or
switching from physics jargon to communication theory jargon, a-posteriori log-likelihood) at the
$n_{\rm it}$-th iteration in the center of the tree one derives, $m^{(n_{\rm it})}_{\rm center} =
{\bm h} \cdot {\bm n}$. $n_i$, defined on a bit of the original Tanner graph is an integer. It is
a sum of contributions, each originating from the respective bit/replica on the computational
tree. Colored are bits on the tree that contribute the integer $+1$ or $-1$. (In the Figures of
the main text and Supporting Figures we use different colors to identify bits in the computational
tree originating from different bits of the original Tanner graph.) A colored bit on the
computational tree has the signature $+1/-1$ if it contributes $+1/-1$ to the integer associated
with the respective bit on the Tanner graph. Uncolored bits (i.e.\ bits not shown in the Figures)
or pale bits (i.e.\ bits shown pale in the Figures) on the tree do not contribute to the
respective integer (one may also say that the respective contribution is just zero) according to
the $\mbox{min}$-part of the min-sum rules described in Eq.~(1) of the main text. We draw a bit on
the computational tree in pale if it does not contribute to respective integer, however at least
one of its siblings, i.e.\ bits on the tree originating from the same bit on the computational
tree, does contribute to the integer.)

Let us now describe how an individual contribution of a colored bit on the tree to the respective
integer (that is $+1$ or $-1$) is calculated. We aim to calculate the contribution to
magnetization at the tree center counting integers according to the min-sum rule. We assign $+1$
signatures to the colored leaves of the tree and start an iterative procedure which assigns
signatures moving from the tree leaves towards the tree center. Consider the case when at a
certain step of the iteration procedure a check receives messages from some number of bits among
which only one is colored. Then one calculates the product of the signatures associated with the
messages this check receives from the remaining bits. If the resulting product is $+1/-1$ the
signature of the colored bit is $+1/-1$ and the signatures of the colored bits, laying on the tree
branch grown from the given colored bit, do not/do change.  The other possibility (that will be
called degenerate) is that a check receives two (or more) messages that all have the same absolute
value, which is also the minimal of all the messages received. Then, one has the freedom to color
only one of the degenerate bits with a colored branch grown from it with the signatures assigned
as described above. The iterative procedure of the signature assignment is terminated once the
three center is reached. One calls a check marked if it lies in between two colored bits of
different signatures.

The union of all colored bits, i.e.\ bits contributing to $m^{(n_{\rm it})}_{\rm center}$, is
called the colored structure. Any check connected to the colored structure is actually connected
to two bits of the structure. Another important characteristic complementing the notion of the
colored structure on the computational tree is the set of aforementioned signatures $\pm 1$
associated with any bit of the colored structure. In a degenerate case, one finds multiple
colored/signature structures associated with a given configuration of magnetic fields. Each
degenerate structure will actually correspond to a distinct linear combination of magnetic fields
equal to $m^{(n_{\rm it})}_{\rm center}$. Therefore, of the whole variety of possible degenerate
colored/signature structures corresponding to the same magnetic field one can always select a set
of linearly independent ones.

So far, this description was generic, i.e.\ not restricted to a specific configuration of the
magnetic field. Let us now fix the family of linear independent colored/signature structures just
explained and allow variation in the value of magnetic fields. Our goal here is to find an
instanton conditioned to the specific form of the family of the linear independent
colored/signature structures. Finding the instanton means minimizing $l^2 = ({\bm 1}-{\bm h})^2$
with respect to ${\bm h}$ under the additional set of linearly independent conditions, $m^{(n_{\rm
it})} = {\bm h} \cdot {\bm n}^{(\mu)} = 0$, where $\mu$ is an index enumerating these conditions
each corresponding to a certain colored structure. (The expression presented above for the length
$l$ applies to the white Gaussian channel,  however generalization for any other channel is
straightforward.) The resulting expression for the optimal configuration of the noise,
${\bm\varphi} = {\bm 1} - {\bm h}$, is
\begin{equation*}
  {\bm\varphi} = \sum_{\mu,\nu} {\bm n}^{(\mu)}
    (\hat{G}^{-1})_{\mu\nu} \sum_i n_i^{(\nu)},\quad
  G_{\mu\nu}\equiv {\bm n}^{(\mu)}\cdot{\bm n}^{(\nu)}.
  \label{lin_h_MS}\tag{S1}
\end{equation*}
This expression (also generalizing Eq.~(3) of the main text for an arbitrary type of instanton
degeneracy) should be checked for consistency with the family of the colored/signature structures
assumed for the instanton. If the consistency check is met, the instanton construction is
completed.

Let us now demonstrate how this formal description works for the three
instanton examples (a), (b)
and (c) described in Figs.~1,~2 of the main text and also illustrated in
Figs.~S3--S6. Instantons
(a) and (c) are both explained by a single colored structure. For the
(a) instanton each bit
contributes $+1$ to the respective component of ${\bm n}$. For the (c)
example all contributions
are $+1$ except of the one coming from the ``75'' bit connected to the marked check. Since $h_{0}
< 0$, the message originating from this bit contributes with the opposite sign to the
magnetization. There exist $6$ replicas of the ``75'' bit on the computational tree, however
taking into account that one replica of the bit contributes $-1$, one finds that the actual value
of the noise is $\sim 4$ rather than $\sim 6$. Since $|h_{0}| > |h_{75}|$, the ``0'' bit is pale
thus it does not contribute to the magnetization. Considering the (b)
instanton one finds that
this is a degenerate example, with $|h_{0}| = |h_{77}|$. There are two linearly independent
colored/signature structures describing the (b) instanton. The two
structures are different only
at the two bits on the tree leaves shown in Fig.~1B of the main text adjusted to the marked check.
The first colored structure does not contain bit ``77" (zero contribution to the magnetization)
while bit ``0'' contributes $+1$ to the respective component of ${\bm n}$. The second structure
does not contain bit ``0" while bit ``77'' contributes $-1$ to the respective component of ${\bm
n}$ (simply because $h_{0} < 0$, thus forcing the respective message to contribute to the
magnetization with the opposite sign).

One natural question to ask about the degenerate case (b) is the
following: can one of the two
colored structures describing the instanton be forming its own non-degenerate instanton?  The
answer is negative. Indeed, the colored/signature structure generates (through the minimization
procedure described above) such a configuration of magnetic fields that will not be consistent with
the colored/signature structure one started from. Considering the structure with the ``0'' bit
connected to the marked check being pale (and thus the signature field associated with the colored
bit ``77'' connected to the red check being $-1$) one finds that the magnetic field minimizing $l$
will actually be inconsistent with the colored/signature structure. Considering the other
configuration (the ``0'' bit is colored with $+1$ signature while the ``77'' bit is pale) one
finds again that the resulting magnetic field is inconsistent with the colored structure. To
resolve this inconsistency one needs to account for the two configurations simultaneously, thus
introducing two constraints, not one. The degeneracy of the (b)
instanton  is illustrated in
Fig.~S6.

Let us also notice that the discrete nature of consistency check (yes/no answer as a result) puts
degenerate configurations on equal footing (in the sense of counting all possibilities) with the
non-degenerate configurations: considering the family of all possible instantons for the given
computational tree one finds that the number of degenerate instantons is comparable with the
number of nondegenerate instantons.

\subsection{Instantons as medians between pseudo-codewords}

We consider an instanton with the set of linearly independent structures already established. Each
colored structure,  indexed by $\mu$, corresponds to the constraint, ${\bm h}\cdot{\bm
n}^{(\mu)}=0$, imposed on the magnetic field,  ${\bm h}$. Each of the constraints can actually be
reformulated in terms of a pair of pseudo-codewords on the computational tree:
\begin{equation*}
\sum\limits_{i\in{\rm tree}} h_{i} \sigma_{i}^{(\mu;+)}=\sum\limits_{i\in{\rm tree}} h_{i}
\sigma_{i}^{(\mu;-)}, \label{pseudo} \tag{S2}
\end{equation*}
where $i$ stands for index assigned to a bit on the computational tree; the magnetic field on the
computational tree bit is equal to the magnetic field defined on the respective bit of the original
graph; and the pseudo-codewords, ${\bm \sigma}^{(\mu;\pm)}$, are the two distinct configurations
of the binary field, $\sigma_i=\pm 1$, defined on each bit $i$ of the computational tree that
satisfy all the checks on the computational tree.

Let us now discuss how the pseudo-codewords can be constructed if the respective constraint $\mu$
described by Eq.~(\ref{pseudo}), is already established. If the signature field, described in the
previous Section of the Notes, corresponding to the structure $\mu$, does not contain a single
$-1$ element, then ${\bm \sigma}^{(\mu;+)}$ is the all unity codeword ($+1$ on all bits of the
computational tree) and ${\bm \sigma}^{(\mu;-)}$ is the pseudo-codeword containing $-1$ on all the
colored bits of the structure and $+1$ on all other bits. If, however, the colored structure does
contain some $-1$ signatures the situation is more elaborate as both pseudo-codewords are
nontrivial. The algorithm that allows restoration of the pseudo-code words starts by determining
the values of the colored bits for ${\bm\sigma}^{(\mu;+)}$ that are set equal to the values of the
signatures. The uncolored bits are assigned values $+1$. Although it is possible to determine the
bit values of the pale substructures the procedure is elaborate and we are not presenting it here.
Finally, the pseudo-codeword ${\bm\sigma}^{(\mu;-)}$ is obtained from ${\bm\sigma}^{(\mu;+)}$ by
changing the signs of colored bits with the uncolored and pale bits remaining the same.

Fig.~2 of the main text shows three examples of the pseudo-codeword construction for the three
instantons discussed in the manuscript. Examples (a), (c) contain one
pair of competing
pseudo-codewords. However, the two cases are different. In the case (a)
the colored/signature
structure does not contain $-1$ bits, thus one pseudo-codeword is just the all unity codeword and
another pseudo-codeword contains $-1$ at all the bits of the colored structure and $+1$ at all
other bits. In the (c) case the colored/signature structure does contain
$-1$ bit thus resulting
in two distinct pseudo-codewords shown in Fig.~2C of the main text.
Example (b) corresponds to the
degenerate case with the two pairs of pseudo-codewords involved in the conditions (\ref{pseudo}).
However, one pseudo-codeword enters both conditions (that is the one shown on the top diagram of
Fig.~2B of the main text) therefore the total count for the case (b)
gives three pseudo-codewords
being equidistant from the instanton configuration of the magnetic fields.

\subsection{General Remarks}

Let us note that the analysis presented above, in addition to its theoretical significance, may be
helpful for accelerating the instanton-amoeba numerical procedure, e.g.\ through guiding selection
on the final stage of the minimization. We also expect that this theoretical analysis will be
instrumental for formulating the right questions to address by the instanton-amoeba method, or by
other minimization methods aiming at finding the instanton numerically.  In what follows, we
conclude by posing some questions that we did not yet study but plan to address in the future.

\begin{itemize}

  \vspace{-2mm}

 \item More detailed exploration of the phase space, especially in the
   context of describing not only the minimal  distance $l_{\rm min}$
   contribution but also the family of other ``low laying'' instantons.
   The particular question of interest here is to estimate the ``density
   of states/instantons'',  that is to answer the question: how many
   instantons are found within the $\delta l$ vicinity of the one
   correspondent to $l_{\rm min}$?

  \vspace{-2mm}

 \item Dependence of BER on the number of iterations. As we already
indicated in the main text our preliminary tests show that instantons and thus asymptotic
estimates for BER do change with the number of iterations. We will be interested to explain this
dependence.  We will also be testing with our instanton-amoeba approach, the validity of the graph
covers method suggested recently \cite{03KV}.

  \vspace{-2mm}

 \item Dependence on the code length. It is important to analyze the
family of LDPC codes with varying code length, $N$. Of a special interest are the regular LDPC
codes where the Hamming distance grows with $N$, e.g.\ Margulis codes \cite{82Mar}. Then, the
relevant question is: how does $l_{\rm min}$ (and other characteristics of the error floor) change
with $N$ for a given family of codes? This study will essentially lead to analysis of the
finite-size effects, already discussed in the water-fall domain \cite{04ARMR}, but not yet
explored in the asymptotic regime of the error-floor.

  \vspace{-2mm}

 \item Does BP/min-sum decoding perform better than other suboptimal
algorithms (that can possibly exist) of the same complexity, e.g.\ linear in $N$? Even if the
answer is yes (that is by no means guaranteed), what would be the best decoding for a higher level
of complexity, e.g.\ $N^a$, where $a>1$? Once an idea of better decoding is formulated, our
instanton-amoeba toolbox will be indispensable in answering the aforementioned questions and also
testing in depth the performance of the new decoding.

  \vspace{-2mm}

 \item
Other types of codes,  e.g.\ turbo codes. Turbo codes show remarkable performance at moderate SNR
but they are also infamous for demonstrating much higher (than comparable in size LDPC codes) error
floors. Even though some important similarities between the LDPC codes and turbo codes are
established \cite{98MMC},  the decoders of these two types of codes are different and it becomes
important to analyze the performance of the turbo scheme, especially in light of the turbo-codes
popularity.

  \vspace{-2mm}

\item Other, application specific, channels. The instanton-amoeba
approach is not limited to the white Gaussian channel,  which we choose primarily for the purpose
of demonstration, but can be applied straightforwardly to other types of channels,  e.g.\ with
correlations among received samples. Of special interest will be to analyze the performance of
fiber-optic communication channels where the effects of fiber dispersion \cite{03CCDGKL},
birefringence and amplifier noise \cite{04CCGKL} will be accounted for. Another two interesting
channel types are magnetic and optical recording channels exhibiting high level of nonlinearity
and correlations among received samples \cite{04BV_book}.

  \vspace{-2mm}

 \item
There are many problems in the information and computer sciences that are different from standard
coding problem but are also dependent or sensitive to rare errors. Therefore, estimating
performance/BER in these problems is a major step required for their comprehensive analysis. Two
interesting examples here are (i) inter-symbol interference, that is especially challenging in the
context of two-dimensional \cite{03WOSI} and three-dimensional information storage,  and (ii)
estimating algorithmic errors in the domain of typically good performance within a combinatorial
optimization K-SAT setting \cite{02MPZ}.

\end{itemize}

\end{document}